\documentclass[runningheads]{llncs}
\usepackage{graphicx}
\usepackage{pgfplots}
\usepackage{tabulary,lipsum}
\usepackage{float}
\usepackage[inline]{enumitem}
\usepackage[nolist,nohyperlinks]{acronym}
\usepackage{todonotes}
\usepackage{csquotes}
\pgfplotsset{compat=newest}
\pgfplotsset{
    box plot/.style={
        /pgfplots/.cd,
        black,
        only marks,
        mark=-,
        mark size=\pgfkeysvalueof{/pgfplots/box plot width},
        /pgfplots/error bars/y dir=plus,
        /pgfplots/error bars/y explicit,
        /pgfplots/table/x index=\pgfkeysvalueof{/pgfplots/box plot x index},
    },
    box plot box/.style={
        /pgfplots/error bars/draw error bar/.code 2 args={%
            \draw  ##1 -- ++(\pgfkeysvalueof{/pgfplots/box plot width},0pt) |- ##2 -- ++(-\pgfkeysvalueof{/pgfplots/box plot width},0pt) |- ##1 -- cycle;
        },
        /pgfplots/table/.cd,
        y index=\pgfkeysvalueof{/pgfplots/box plot box top index},
        y error expr={
            \thisrowno{\pgfkeysvalueof{/pgfplots/box plot box bottom index}}
            - \thisrowno{\pgfkeysvalueof{/pgfplots/box plot box top index}}
        },
        /pgfplots/box plot
    },
    box plot top whisker/.style={
        /pgfplots/error bars/draw error bar/.code 2 args={%
            \pgfkeysgetvalue{/pgfplots/error bars/error mark}%
            {\pgfplotserrorbarsmark}%
            \pgfkeysgetvalue{/pgfplots/error bars/error mark options}%
            {\pgfplotserrorbarsmarkopts}%
            \path ##1 -- ##2;
        },
        /pgfplots/table/.cd,
        y index=\pgfkeysvalueof{/pgfplots/box plot whisker top index},
        y error expr={
            \thisrowno{\pgfkeysvalueof{/pgfplots/box plot box top index}}
            - \thisrowno{\pgfkeysvalueof{/pgfplots/box plot whisker top index}}
        },
        /pgfplots/box plot
    },
    box plot bottom whisker/.style={
        /pgfplots/error bars/draw error bar/.code 2 args={%
            \pgfkeysgetvalue{/pgfplots/error bars/error mark}%
            {\pgfplotserrorbarsmark}%
            \pgfkeysgetvalue{/pgfplots/error bars/error mark options}%
            {\pgfplotserrorbarsmarkopts}%
            \path ##1 -- ##2;
        },
        /pgfplots/table/.cd,
        y index=\pgfkeysvalueof{/pgfplots/box plot whisker bottom index},
        y error expr={
            \thisrowno{\pgfkeysvalueof{/pgfplots/box plot box bottom index}}
            - \thisrowno{\pgfkeysvalueof{/pgfplots/box plot whisker bottom index}}
        },
        /pgfplots/box plot
    },
    box plot median/.style={
        /pgfplots/box plot,
        /pgfplots/table/y index=\pgfkeysvalueof{/pgfplots/box plot median index}
    },
    box plot width/.initial=1em,
    box plot x index/.initial=0,
    box plot median index/.initial=1,
    box plot box top index/.initial=2,
    box plot box bottom index/.initial=3,
    box plot whisker top index/.initial=4,
    box plot whisker bottom index/.initial=5,
}
\usepackage[style=lncs,giveninits=true,maxnames=3,mincitenames=1,maxcitenames=1,backend=biber]{biblatex}
\usepackage{tabulary,lipsum}
\usepackage{tabularx}
\usepackage{microtype}
\usepackage{booktabs}

\usepackage{xurl}
\usepackage[hidelinks,bookmarks=false,final,breaklinks]{hyperref}
\usepackage{listings}
\usepackage[lighttt]{lmodern}
\usepackage[absolute]{textpos}
\lstdefinestyle{dsl}{
    morekeywords={target,import,type,using,data,node,identity,attribute,property,name,class,constraint,element},
    emph={NEVER,FLOWS,WHERE},
    emphstyle={\textit}
}

\lstdefinestyle{ddc}{
    morekeywords={enum, using, enumCharacteristicType}
}

\lstdefinestyle{java}{
    language=Java,
    morekeywords={var,if,return},
}

\newcommand{\lstbegin}{\begin{minipage}{0.94\linewidth}\bigskip\medskip}
\newcommand{\lstend}{\medskip\end{minipage}}

\begin{document}

\begin{textblock}{14}(1,0.5)
\noindent \textbf{This is a preprint of the paper submitted and accepted at ECSA Tool \& Demo Track 2023. For this track, there will be no formal proceedings. An extended version might be published in the Postproceedings by Springer LNCS.} 
\end{textblock}

\title{Tool-Supported Architecture-Based Data Flow Analysis for Confidentiality
}

\titlerunning{Data Flow Analysis for Confidentiality}

\author{Felix Schwickerath \inst{1,2} \and
Nicolas Boltz\inst{1,3} \and
Sebastian Hahner\inst{1,3} \and
\\Maximilian Walter\inst{1,3} \and
Christopher Gerking \inst{1,3} \and
Robert Heinrich \inst{1,3}
}
\authorrunning{F. Schwickerath et al.}

\institute{Karlsruhe Institute for Technology (KIT) \and \email{felix.schwickerath@student.kit.edu} \and
\email{\{boltz,hahner,maximilian.walter,gerking,heinrich\}@kit.edu}}

\maketitle

\begin{acronym}
	\acro{pcm}[PCM]{Palladio Component Model}
	\acro{add}[ADD]{Architectural Design Decision}
	\acro{dsl}[DSL]{Domain Specific Language}
	\acro{dfd}[DFD]{Data Flow Diagrams}
    \acro{seff}[SEFF]{Service Effect Specification}
\end{acronym}

\begin{abstract}
Through the increasing interconnection between various systems, the need for confidential systems is increasing. Confidential systems share data only with authorized entities.
However, estimating the confidentiality of a system is complex, and adjusting an already deployed software is costly.
Thus, it is helpful to have confidentiality analyses, which can estimate the confidentiality already at design time. Based on an existing data-flow-based confidentiality analysis concept, we reimplemented a data flow analysis as a Java-based tool. The tool uses the software architecture to identify access violations based on the data flow. The evaluation for our tool indicates that we can analyze similar scenarios and scale for certain scenarios better than the existing analysis.

\keywords{Confidentiality  \and Software Architecture \and Security.}
\end{abstract}


\section{Introduction}
\label{sec:Introduction}
With increased digitalization, more and more systems and digital services are integrated into our lives. These systems often gather data to enable efficient services, like a purchase history in an online shop. This collected data is then exchanged with other services or systems. For instance, in the case of an online shop, customer data might be shared with payment providers. Often, the collected data contains sensitive data, such as the mentioned payment information or a customer's address. Therefore, there is a need to preserve the data's confidentiality.

Confidentiality is described by ISO 27000 as the property ``that information is not made available or disclosed to unauthorized individuals, entities, or processes''~\cite[Section 3.10]{isoConfidentiality}. A system violating confidentiality can result in privacy violations, which can result in costly fines, as seen in the case of H\&M~\cite{the_hamburg_commissioner_for_353_2020} or British Airways~\cite{british_fine}. However, identifying confidentiality violations can be difficult, because the connected services build a complex network of data flows. Hence, a systematic approach to analyze them is required.


Data flow analyses based on source code, e.g., JOANA \cite{Joana} or KeY \cite{key}, cannot consider context information, such as deployment. However, deployment information can be essential for confidentiality, because the deployment can contain whether the application is deployed on an external cloud provider or not. In addition, source code analyses cannot be used in early design phases because 
of their need for existing source code. Analyzing the system early at  design time is beneficial, because fixing issues in later phases is usually more costly \cite{shull_what_2002}. Seifermann et al. \cite{PalladioIntegrationConfidentiality, seifermannJSS} proposed an architecture-based data flow analysis to analyze systems for confidentiality violations. The approach can consider additional context information, such as the deployment, enabling software architects to analyze confidentiality during early design phases. 
However, their Prolog-based implementation of the analysis is very hard to maintain and has a high resource (memory) demand, which severely limits the applicability for large systems. Hence, we decided to reimplement the analysis as a  Java-based open-source Eclipse plugin\footnote{Video demonstration available: \url{https://www.youtube.com/watch?v=q3WJsMyqJcA}}.

The approach of Seifermann et al. \cite{PalladioIntegrationConfidentiality, seifermannJSS} consists of a metamodel and an analysis. We explain the metamodel and the scientific concept for the analysis in \autoref{sec:modelling}. In \autoref{sec:Architecture}, we describe the reasons for the reimplementation and our expected benefits. In addition, we give insight into the tool architecture and how it relates to the scientific concept. \autoref{sec:Application} explains how our developed tool can be used. We compare the old analysis with our newly developed tool in \autoref{sec:Evaluation}. For the investigated scenarios, our comparison shows that we can identify the same violations, and we need fewer resources to analyze larger systems. In the last section, we conclude the paper and discuss future work. 

\section{Modeling Confidentiality in Software-Architectures}
\label{sec:modelling}

Our analysis approach uses software-architectural models to determine the confidentiality of a software system.
Here, we build on the Palladio Component Model (PCM) \cite{reussner2016a} as Architectural Modeling Language (ADL).
Using PCM is beneficial, since it supports security analyses \cite{hahnerSeams, WalterTargeted, walter_architectural_2022} as well as performance and reliability analyses \cite{reussner2016a}, thereby reducing the overall effort required by software architects. 
PCM was a foundation in the original data flow analysis \cite{seifermannJSS}.

\begin{figure}
    \centering
    \setlength{\abovecaptionskip}{3pt}
    \includegraphics[width=0.8\textwidth]{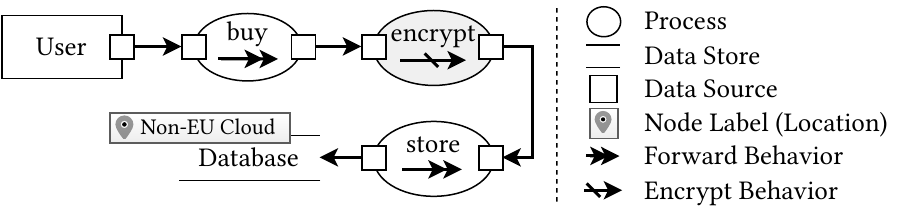}
    \caption{Data flow diagram that represents the flow of user data in the running example}
    \label{fig:DFD}
\end{figure}

To enhance the description of our modeling and analysis, we present the running example of an online shop that is deployed within the European Union~\cite{Hahner_2021}.
Using this online shop, users can browse through the available inventory of items and select an item to purchase.
Here, sensitive information, like the user's address, is sent to the online shop, which encrypts this data and stores it in a database that is deployed outside the EU, as shown in \autoref{fig:DFD}.
Without encryption, this data flow would violate confidentiality.

PCM enables us to describe the software architecture of the online shop from different viewpoints.
The structure of the software architecture is modeled as multiple components, e.g., representing the shop interface and the database, and connected in the assembly model.
The behavior of the system, e.g., calling of services and processing of data, is modeled as \emph{ServiceEffectSpecifications (SEFF)}.
User behavior is captured in the usage model, which contains multiple usage scenarios, each describing the service calls by a user.
Lastly, the hardware of the system is represented in the resource environment, and deployment information is stored in the allocation model.

Seifermann et al. \cite{PalladioIntegrationConfidentiality, Seifermann2022, Seifermann2019,Seifermann_Walter_Hahner_Heinrich_Reussner_2021} extended PCM to annotate confidentiality-related properties like data sensitivity or encryption and automatically derive a data flow diagram from the architecture. 
Afterward, the diagram is analyzed in a data flow-based confidentiality analysis \cite{seifermannJSS}. 
The concept was reused by different architectural analyses targeting uncertainty \cite{boltz_handling_2022,hahnerClassificationUncertainty,Hahner2023Uncertainty,WalterPeropteryx}, Industrial IoT \cite{boltz_context-based_2020} or estimating attacker impacts \cite{walterAttackATDataflow}.

In the remainder of this section, we briefly describe the concept of the data flow analysis using the running example shown in \autoref{fig:DFD}.
The automatically derived data flow diagram contains confidentiality-related information that has been extracted from the annotated software architecture model.
This includes the behavior of nodes, e.g., encrypting or only forwarding data.
Data labels represent the characteristics of the data within the system, e.g., whether the data is encrypted.
Node labels represent characteristics of the system itself, e.g., the non-EU deployment location of the database component.
All available characteristics are listed in a data dictionary and can be used to define data flow constraints \cite{Hahner2021b}.
In our running example, we restrict user data labeled as \emph{personal}, but not labeled as \emph{encrypted}, from flowing to a \emph{non-EU} labeled node.

The data flow analysis checks these constraints using \emph{label propagation} \cite{seifermannJSS}.
Data flows through the data flow diagram and can be altered by the nodes' behavior, e.g., by adding the \emph{encrypted} label.
In each node, the constraints are examined, taking into account all propagated data labels and also the node's label.
In our running example shown in \autoref{fig:DFD}, a constraint violation would occur if we remove the \emph{encrypt} node, which is highlighted gray.
In this case, the \emph{personal} label would propagate to a \emph{non-EU}-labeled node without the \emph{encrypted} label, which violates confidentiality.
This analysis was originally implemented using Prolog.
By transforming the data flow diagrams and all of their properties into facts, the Prolog environment can solve queries.
Architects can either define their constraints directly in Prolog or by using a domain-specific language \cite{Hahner2021b}.

\section{Analysis Architecture}
\label{sec:Architecture}
\begin{figure}
    \centering
    \setlength{\abovecaptionskip}{3pt}
    \setlength{\belowcaptionskip}{-3pt}
    \includegraphics[width=\textwidth]{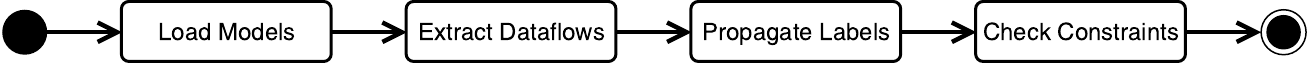}
    \caption{Analysis architecture as performed activities.}
    \label{fig:ImplStructure}
\end{figure}

The data flow analysis by \cite{seifermannJSS} as described in \autoref{sec:modelling} is made up of four steps. \autoref{fig:ImplStructure} shows the analysis steps and their sequential order as an activity diagram.
First, the \ac{pcm} and analysis-specific models are loaded and references between model elements are resolved. This is done automatically by EMF.
Using the information from the models and annotations, described in \autoref{sec:modelling}, possible data flows are extracted. As \ac{pcm} allows developers to model different use cases, the analysis first needs to determine all possible starting points of data flows. For each starting point, the analysis iterates the following calls and adds a node to the data flow for each encountered element. Calls to \acp{seff} defined in interfaces are handled differently:
For each call encountered, the analysis adds a calling and returning node in the data flow, as returning values from \acp{seff} are allowed.%

After all data flows are extracted, the analysis propagates the characteristic labels that are defined in the data dictionary model and have been added to the \ac{pcm} models using annotations. Starting at the first node in the data flow, the analysis evaluates the node characteristics that are present at the given node. Using the node characteristics applicable at the current node and the node characteristics from the previous node, the analysis is able to resolve the defined relationship between inputs/pins and the characteristics of the node.
Furthermore, as data characteristics are applied to variables and parameters, the analysis filters the variables with their data characteristics to only include variables that are in scope.%

Using the data flows and propagated characteristic labels, data flow constraints can be checked. For example, by comparing propagated data characteristics with defined node characteristics, as described in \autoref{sec:modelling}.

The Prolog-based analysis of \cite{seifermannJSS} realized the extraction of data flows and propagation of labels, by first transforming the \ac{pcm} models to an explicit DFD metamodel notation, then transforming the DFD elements to Prolog statements and rules. Data flow constraints are checked by defining Prolog queries that are unique to the modeled system and defined data dictionary model. As one DFD element with characteristics is transformed into multiple Prolog statements, the Prolog code grows exponentially with the model size. The exponential growth results in high demand of memory, as the whole Prolog program needs to be fully loaded by the Prolog interpreter.

Additionally, the formulation of constraints and the debugging of issues can become complex due to Prolog.
A DSL, as proposed by \textcite{Hahner2021b}, can help users to formulate queries to the Prolog model, but the added step of indirection makes it even more difficult to extend the Prolog-based analysis.
As the analysis is made up of multiple chained transformations and intermediate model representations, the maintenance of the analysis is made even harder. 


Due to the aforementioned reasons, our reimplementation realizes all steps of the analysis using Java. Our reimplementation is based on the current \ac{pcm} version and does not consider plugins labeled as \emph{incubation}. We extract data flows and represent them in simple ordered lists called \texttt{Action\-Sequence}. An \texttt{Action\-Sequence} is made up of \texttt{Action\-Sequence\-Elements}, each representing a node in a data flow. We propagate the characteristic labels for each \texttt{Action\-Sequence} individually, by iterating the contained elements and saving the result of the propagation for each node in the corresponding \texttt{Action\-Sequence\-Element}. Doing so not only eliminates the requirement of using logical programming languages, but also removes the need for both transformations and intermediate model representations of the Prolog-based analysis. Thereby, we drastically simplify and reduce maintenance effort. We also create \texttt{Action\-Sequences} with immutable elements, ensuring that no data is shared between \texttt{Action\-Sequences}. This separation of \texttt{Action\-Sequences} allows for the parallelization of the extraction of data flows, propagation of labels, and evaluation of constraints in the future. 

Additionally, data and node characteristics are propagated independently of the constraint of the analysis. Due to this reason, our reimplementation is able to evaluate multiple constraints without propagating characteristic labels again. Compared to the Prolog-based analysis, this drastically improves the performance when analyzing a system model for multiple constraints.

\section{Tool Application}
\label{sec:Application}

The Java-based re-implementation of the data flow analysis 
is available as open source tool based on Eclipse Ecore and the Eclipse Modeling Tools.
Documentation and installation guidance of our tool can be found in our repository \cite{github}. 
We also provide example models that are used as test models to ensure the analysis produces correct results compared to the Prolog-based analysis.
\autoref{lst:codesnippet} demonstrates the usage of the analysis using the running example.
We provide a \emph{builder} to set up the analysis with required inputs, which is simplified in line~1.
After initializing the analysis in line~2, all possible data flows, i.e., sequences, are extracted from the architectural model in line~3.
In line~4, we propagate all annotated labels through these data flows.
After the label propagation, we search for constraint violations starting in line~6.
For each possible data flow in the modeled software architecture, we test each data flow node for a predicate that represents the constraint.

\begin{lstlisting}[
    float,
    style=java,
    caption={Code snippet showing how to initialize and how to use the analysis},
    abovecaptionskip=8pt,
    belowcaptionskip=-3pt,
    label={lst:codesnippet}
]
  var analysis = new DataFlowAnalysisBuilder().build(); // simplified
  analysis.initializeAnalysis();
  var allSequences = analysis.findAllSequences();
  var propagationResult = analysis.evaluateDataFlows(allSequences);
      
  for (var sequence : propagationResult) {
    var violations = analysis.queryDataFlow(sequence, node -> {
      if (node.hasNodeCharacteristic("ServerLocation", "nonEU")) {
        return node.getAllDataFlowVariables().stream().anyMatch(v -> 
          v.hasDataCharacteristic("DataSensitivity", "Personal") && 
          !v.hasDataCharacteristic("Encryption", "Encrypted"));
      }
      return false;
    });
  }
\end{lstlisting}

In our running example, we only look for nodes that are located outside the EU in line~8.
In lines 9 to 11, we evaluate whether one of these nodes receives \emph{personal} data that has no \emph{encrypted} label.
If this is the case, for any node in any possible data flow, the data flow constraint is violated and the violation is collected.
After the execution, the variable \emph{violations} contains a list of all constraint-violating nodes within the modeled software system.
If no violation has been found, the list remains empty.

\section{Evaluation}
\label{sec:Evaluation}
For our evaluation, we aim to compare our Java-based analysis to the Prolog-based analysis.
Our evaluation goals are to examine the accuracy and scalability of both analyses, to show that our reimplementation retains the core functionality of the Prolog-based analysis, while improving execution times and resource demand.

\subsection{Evaluation Design}
To examine and compare accuracy, we check whether both analyses are able to correctly identify violations, using various 
\ac{pcm} models. 
To ensure a good base for comparison, we reuse the case study-based models that were originally used by \textcite{seifermannJSS} to evaluate the accuracy of the Prolog-based analysis.
We selected the case studies using the default call return semantics of the current stable version of \ac{pcm}.
As a metric, we count correctly identified violations. 

To examine and compare scalability, we check the full execution time of both analyses, when analyzing models with increasing size. To better distinguish the impact of different features of the models on the scalability, we generate individual minimal models with an increasing number of either node characteristics, characteristic label propagation, variable actions or SEFF parameters. We chose these elements, as they have the highest impact on either the length of Prolog code or Java loop iterations, depending on the analysis. For each run, we increase the model feature under consideration by the power of ten, starting at $10^0$ and ending with $10^5$.
Each analysis is run with a constraint, which finds a violation at each node, forcing each node to be evaluated once. The constraint ensures a worst-case execution time for both analyses. We run each test 10 times and take the median execution time to exclude outliers or measurement anomalies.
We executed the analyses on a dedicated VM. The VM has 4 AMD Opteron 8435 cores together with 97 GB RAM and runs Debian 11 with OpenJDK 11/17. 

\subsection{Evaluation Results}
\label{sec:EvaluationResults}
Regarding accuracy, both analyses were able to correctly identify the 42 violations that were present in the case study-based models and did not return any false positives. \autoref{tab:Validation} shows the results of the accuracy evaluation and size of analyzed models. For a better overview, the results have been aggregated based on the underlying case study that has been analyzed. As both analyses performed the same, we assume, that our reimplemented Java-based analysis is functionally equivalent to the Prolog-based analysis, when analyzing models using the call return semantics.
\begin{table}
    \centering
    \setlength{\abovecaptionskip}{8pt}
    \setlength{\belowcaptionskip}{-5pt}
    \begin{tabularx}{\textwidth}{l>{\centering\arraybackslash}X >{\centering\arraybackslash}X|cc} 
        \toprule
        Case Study & Prolog-based & Java-based & Components & Labels \\
        \midrule
        ContactSMS \cite{Katkalov2017} & 10 violations & 10 violations & 3 & 4 \\
        FlightControl \cite{Seifermann2022Detecting} & 0 violations & 0 violations & 6 & 6\\
        FriendMap \cite{Tuma2019} & 0 violations & 0 violations & 5 & 12 \\
        Hospital \cite{Tuma2019} & 0 violations & 0 violations & 4 & 12 \\
        ImageSharing \cite{Seifermann2022Detecting} & 0 violations & 0 violations & 1 & 9 \\
        PrivateTaxi \cite{Katkalov2017}  & 0 violations & 0 violations & 13 & 20 \\
        TravelPlanner \cite{Katkalov2017} & 32 violations & 32 violations & 7 & 8 \\
        WebRTC \cite{Tuma2019} & 0 violations & 0 violations & 20 & 12 \\
        \bottomrule
    \end{tabularx}
    \caption{Results of both analyses compared and size of case study-based models}
    \label{tab:Validation}
\end{table}

Regarding scalability, we plotted the results of both analyses as line graphs for each examined model feature, shown in \autoref{fig:graph}. Each graph contains data points from both, the Prolog-based analysis, colored red, and the Java-based analysis, colored blue. Both axes have logarithmic scaling. The x-axis shows the increasing number of model elements and the y-axis the median of execution times in milliseconds.
Our evaluation showed that the Prolog-based analysis is not able to complete a run for more than 1000, for node characteristics, or 100, for variable actions and SEFF parameters. As described in \autoref{sec:Architecture} the Prolog-based analysis has a high demand in system memory. In our tests, the analysis ran into \emph{out of memory errors} or crashed, despite the 97 GB of memory. 

When increasing the number of characteristic propagation, the execution time behavior of both analyses is similar. However, for the other evaluated cases, we can observe, that our reimplemented Java-based analysis retains a nearly constant execution time up to $10^3$ elements, while the Prolog-based analysis shows an at least linear increase in execution times or fails to complete the analysis run.

The exponential increase in execution time of the Java-based analysis for larger models can be explained by some inefficiencies in sequence finding, overhead during characteristics propagation, and tradeoffs of our immutable approach to action sequences.
Nonetheless, we reckon that the time required in all cases is still feasible for design-time analyses. Compared to the Prolog-based analysis, the feasible execution times and ability to even analyze large models make our reimplementation more usable for real-world systems. To overcome the lack of replication packages 
\cite{Konersmann}, we provide a data set \cite{dataset}.

\usepgfplotslibrary{groupplots}
\usetikzlibrary{matrix}

\begin{figure}[h]
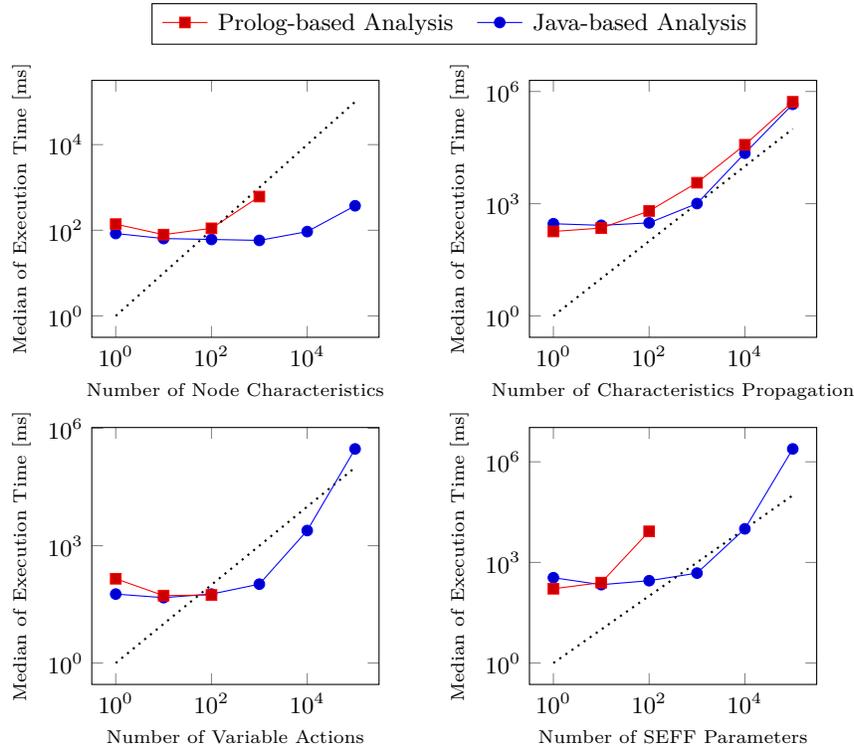

    \centering
    \setlength{\abovecaptionskip}{3pt}
\begin{tikzpicture}
    \begin{groupplot}[group style={group name = graphs,group size = 2 by 4, horizontal sep=2cm, vertical sep=1.2cm}, height=5cm, width=5.4cm]
        \input{graphs/NewNodeCharacteristics}
        \input{graphs/NewCharacteristicsPropagation}
        \input{graphs/VariableActions}
        \input{graphs/SEFFParameter}
        \coordinate (bot) at (rel axis cs:1,0);
    \end{groupplot}
    \path (graphs c1r1.north west|-current bounding box.north)--
      coordinate(legendpos)
      (graphs c2r1.north east|-current bounding box.north);
    \matrix[
        matrix of nodes,
        anchor=south,
        draw,
        inner sep=0.2em,
        draw
    ]at([yshift=1ex]legendpos) 
    {
        \ref{plots:OldAnalysis}& Prolog-based Analysis&[5pt]
        \ref{plots:NewAnalysis}& Java-based Analysis\\};
\end{tikzpicture}
    \caption{Scalability Results Prolog-based Analysis and Java-based Analysis}
    \label{fig:graph}
\end{figure}

\section{Conclusion}
\label{sec:Conclusion}
In this paper, we showcase our Java-based reimplementation of a data flow analysis, based on the approach and tooling of \textcite{seifermannJSS}. Related approaches and tools are described in the previous publications \cite{seifermannJSS, Seifermann2019}. We show how to model confidentiality in software architecture and describe the abstract architecture of the analysis. We highlight problems of the Prolog-based analysis of \textcite{seifermannJSS}, including poor maintainability due to complexity and high demand in system memory, and describe the benefits of our Java-based analysis. Further, we show how to apply our new tooling and evaluate our Java-based analysis by comparing it to the existing Prolog-based analysis. In our evaluation, we show that our reimplemented Java-based analysis is functionally equivalent to the Prolog-based analysis and is able to analyze larger system models.

In future work, we aim to apply our tool 
to constraints regarding privacy as part of a framework for simplified collaboration in legal data protection assessments \cite{boltz2022model}.
We are also currently working to allow explicitly modeled data flow diagram system representations as input. 

\subsubsection*{Acknowledgements}
This publication is partially based on the research project SofDCar (19S21002), which is funded by the German Federal Ministry for Economic Affairs and Climate Action. This work was also supported by funding from the topic Engineering Secure Systems of the Helmholtz Association (HGF) and by KASTEL Security Research Labs, the DFG (German Research Foundation) project number 432576552 (FluidTrust), the BMBF (German Federal Ministry of Education and Research) grant number 16KISA086 (ANYMOS) and "Kerninformatik am KIT (KiKIT)" funded by the Helmholtz Association (HGF).


\printbibliography[heading=bibintoc]
\end{document}